\documentclass[a4paper,fleqn]{cas-dc}

\usepackage[authoryear]{natbib}

\def\tsc#1{\csdef{#1}{\textsc{\lowercase{#1}}\xspace}}
\tsc{WGM}
\tsc{QE}
\tsc{EP}
\tsc{PMS}
\tsc{BEC}
\tsc{DE}

\begin{document}
\let\WriteBookmarks\relax
\def\floatpagepagefraction{1}
\def\textpagefraction{.001}

\shorttitle{Prostate Gland Segmentation with PSHop}

\shortauthors{Yijing Yang et~al.}

\title [mode = title]{PSHop: A Lightweight Feed-Forward Method for 3D Prostate Gland Segmentation}                      




%
\author[1]{Yijing Yang}[]
\author[1]{Vasileios Magoulianitis}[orcid=0000-0001-7511-2910]
\cormark[1]



\ead{magoulia@usc.edu}



\affiliation[1]{organization={Electrical and Computer Engineering Department, University of Southern California (USC)},
    addressline={3740 McClintock Ave.}, 
    city={Los Angeles},
    postcode={90089}, 
    state={CA},
    country={USA}}

\affiliation[2]{organization={Department of Urology, Keck School of Medicine, University of Southern California (USC)},
    addressline={1975 Zonal Ave.}, 
    city={Los Angeles},
    postcode={90033}, 
    state={CA},
    country={USA}}

\affiliation[3]{organization={Department of Radiology, Keck School of Medicine, University of Southern California (USC)},
    addressline={1975 Zonal Ave.}, 
    city={Los Angeles},
    postcode={90033}, 
    state={CA},
    country={USA}}


\author[1]{Jiaxin Yang}[]
\author[1]{Jintang Xue}[]
\author[2]{Masatomo Kaneko}[orcid=0000-0002-1205-807X]
\author[2]{Giovanni Cacciamani}[orcid=0000-0002-8892-5539]
\author[1]{Andre Abreu}[orcid=0000-0002-9167-2587]
\author[3,2]{Vinay Duddalwar}[orcid=0000-0002-4808-5715]
\author[1]{C.-C. Jay Kuo}[]
\author[2]{Inderbir S. Gill}[orcid=0000-0002-5113-7846]
\author[1]{Chrysostomos Nikias}[]

\cortext[cor1]{The corresponding author is with the Electrical Engineering Department of University of Southern California (USC), Los Angeles, USA.}



\begin{abstract}
Automatic prostate segmentation is an important step in computer-aided
diagnosis of prostate cancer and treatment planning. Existing methods of
prostate segmentation are based on deep learning models which have 
a large size and lack of transparency which is essential for physicians. In this
paper, a new data-driven 3D prostate segmentation method on MRI
is proposed, named PSHop. Different from deep learning based methods,
the core methodology of PSHop is a feed-forward encoder-decoder system
based on successive subspace learning (SSL). It consists of two modules: 1)
encoder: fine to coarse unsupervised representation learning with
cascaded VoxelHop units, 2) decoder: coarse to fine segmentation
prediction with voxel-wise classification and local refinement.
Experiments are conducted on the publicly available ISBI-2013 dataset,
as well as on a larger private one. Experimental
analysis shows that our proposed PSHop is effective, robust and
lightweight in the tasks
of prostate gland and zonal segmentation, achieving a Dice Similarity Coefficient (DSC) of $0.873$ for the gland segmentation task. PSHop achieves a competitive
performance comparatively to other deep learning methods, while keeping the model
size and inference complexity an order of magnitude smaller. 
\end{abstract}




\maketitle

\section{Introduction}\label{sec:introduction}

Prostate cancer (PCa) is reported as the second most frequent cancer among men in 2020, with an estimated of almost 1.4 million new cases and 375,000 deaths worldwide~\cite{sung2021global}. In 112 out of 185 countries of the world, it is even the most frequently diagnosed cancer in men. 
International guidelines recommend systematic 12-core transrectal ultrasound-guided biopsy (TRUSGB) in biopsy-naïve men with elevated prostate-specific antigen (PSA) serum levels of $>$3 ng/ml ~\cite{heidenreich2014eau}. Compared to TRUSGB with limitations in the diagnosis, multiparametric magnetic resonance imaging (mpMRI) has been reported to reduce the detection of insignificant prostate cancer~\cite{venderink2018results,stabile2018mri,van2019head}. It has become the imaging method that is best able to detect clinically significant prostate cancer and guide biopsies~\cite{turkbey2016multiparametric}, due to the superior resolution and contrast of imaging, without harming the human body~\cite{wang2012imaging}. 


Based on prostate MRI, prostate segmentation is an important step in the PCa diagnosis and treatment planning with various aims, such as localizing prostate boundaries for radiotherapy~\cite{pasquier2007automatic}, automating the calculation of the prostate volume --key task for the Prostate-Specific Antigen Density (PSA-D) calculation--  as well as track disease progression~\cite{toth2011accurate}, or localizing the region of interest at the beginning of the computer-aided diagnosis (CADx) of PCa~\cite{vos2012automatic,tiwari2013multi,mehta2021computer}. Moreover, zonal segmentation may enhance the PCa detection models \cite{yuan2022z}, since the prostate zones have different visual features \cite{ginsburg2017radiomic} and potentially zonal segmentation can be used to separate the feature extraction process. 

However, manual MR segmentation takes extended time and labor. Radiologists need to mark slice by slice through visual inspection with a high demand of skills and expertise for accurate segmentation. It also comes with high intra- and inter-observer variation. Thus, automated MR prostate segmentation is needed to help improve the accuracy and efficiency in this routinely applied task from radiologists. Due to the various size and shape of the prostate gland across different patients, low contrast between the gland and adjacent structures, imaging artifacts, as well as heterogeneity in signal intensity around endorectal coils (ERCs)~\cite{ghose2012survey, jia2018atlas, jia20193d}, it is still a challenging task.

Existing prostate segmentation algorithms can be categorized into two classes: traditional and deep learning based methods. The traditional methods mainly include contour based \cite{salimi2018fully, ding2003prostate}, atlas based \cite{klein2008automatic,litjens2012multi,dowling2011fast}, deformable models \cite{cootes1993use,knoll1999outlining}, and machine learning based models such as c-means clustering \cite{rundo2017automated, rundo2018fully} and classification \cite{allen2006differential, litjens2012pattern}.

With the development of convolutional neural networks (CNNs), deep learning based segmentation methods~\cite{tian2018psnet,jia20193d,aldoj2020automatic,Liu2020pyramid,sun2020saunet} are pioneered by the popular U-Net \cite{unet} and V-Net \cite{vnet} architectures. They have achieved superior segmentation performance, comparable to that of expert radiologists \cite{isaksson2023comparison}.

Recently, a novel representation learning framework, named Green Learning (GL) was proposed by \cite{kuo2022green}. It provides a linear feature extraction model for image analysis, with key benefits a lightweight model size, as well as transparent and explainable pipeline. GL is based on the Successive Subspace Learning (SSL) methodology for feature extraction that was proposed in a sequence of papers~\cite{kuo2016understanding,kuo2017cnn,kuo2018data,kuo2019interpretable}. These features have multi-scale properties and are extracted in an unsupervised feed-forward manner using core signal processing operations \cite{chen2020pixelhop++}. GL has been applied for medical image analysis in the tasks of Amyotrophic lateral sclerosis (ALS) classification and cardiac MRI segmentation ~\cite{liu2021voxelhop,liu2021segmentation}.

In this work, we propose a novel prostate segmentation method named PSHop, which is built upon the methodology of SSL. The main contributions can be summarized in three folds:
\begin{enumerate}
\item To the best of our knowledge, PSHop is the first work which applies the Green Learning paradigm and SSL methodology on a 3D medical image segmentation problem, thus exploring a new direction in this field.
\item We propose a feed-forward encoder-decoder network for 3D medical image segmentation without the use of back-propagation that makes the entire pipeline transparent. 
\item Our proposed PSHop method is lightweight which significantly reduces the model size compared to other deep learning based methods while keeping comparable performance.
\end{enumerate}

The rest of the paper is organized as follows. Related work is reviewed in Section~\ref{sec:related_work}. The proposed PSHop method is presented in Section~\ref{sec:methods}. The experimental setup and results are then discussed in Section~\ref{sec:experiments_setup} and Section~\ref{sec:results}. Conclusions are finally drawn in Section~\ref{sec:conclusion}.

\section{Related Work}\label{sec:related_work}

\subsection{Traditional methods}
Traditional methods mainly focus on atlas-based, deformable models, and graph cuts optimization. Atlas-based methods \cite{klein2008automatic, litjens2012multi, dowling2011fast} improve the segmentation accuracy by measuring the similarity between target image and multiple atlases. \cite{klein2008automatic} build two deformed atlas images for comparisons, stressing how important the atlas selected images are. For similarity measure they use the normalized mutual information (NMI) and a majority voting algorithm to combine multiple image segmentations. In a work of \cite{litjens2012multi}, a selective and iterative method for performance level estimation is proposed instead of majority voting.

For methods using graph cuts (GC) ~\cite{boykov2001interactive,boykov2001fast}, optimization is applied to find an optimal solution that separates different regions. Images are represented as graphs and segmentation is viewed as a graph cutting optimization process under certain constraints. For example, \cite{chen2012medical} combine graph cuts with active appearance models to improve the segmentation accuracy. Another work \cite{qiu2014dual} solves the dual problem using convex optimization and specifically employs the flow-maximization algorithms in graphs. \cite{tian2017supervoxel} propose supervoxel-based graph cuts and a 3D active contour model for segmentation refinement.

\subsection{Learning based methods}
With the increasing number of accessible data with ground truth annotations, learning based medical image segmentation methods have made a remarkable progress with the support of machine learning and deep learning. At first, fully convolutional network (FCN) was proposed by modifying the existing classification CNN for the segmentation task. Several FCN-based methods for medical image segmentation are then proposed~\cite{unet,vnet,peng2017large,tian2018psnet}. For example, U-Net \cite{unet} and V-Net \cite{vnet} are two representative pioneer work for the 2D and 3D medical image segmentation. In U-Net, combines a contracting path of multiple convolutional layers and an expansive path of up-convolutional layers as an encoder-decoder network struction. The skip architecture in U-Net uses a simply concatenation operation that builds a bridge between encoder and decoder that take advantage of both coarse and fine features. Different from U-Net which processes 2D images, V-Net was generalized to 3D medical image segmentation based on a volumetric processing. It also introduced a novel loss layer based on the Dice coefficient. \cite{peng2017large} proposed a 3D-GCN framework based on FCN where a Global Convolutional Network (GCN) was designed to address both the localization and classification for segmentation. A boundary refinement block was also proposed which models the boundary alignment as a residual structure.

With the promising performance of U-Net and V-Net, more algorithms are proposed having them as the backbone model in recent years. For example, DenseNet-like U-Net \cite{aldoj2020automatic} took advantage of the strengths of both DenseNet and U-Net for segmentation of the prostate gland and its zones. \cite{cciccek20163d} introduced a network for volumetric segmentation that learns from sparsely annotated volumetric images by extending the previous U-Net architecture by replacing all 2D operations with their 3D counterparts. \cite{sun2020saunet} proposed a shape attentive U-Net (SAUNet) where both the texture information and shape information are used to learn the segmentation. \cite{jin20213d} propose a bicubic intepolation for extracting the low frequencies of input MRI as a preprocessing step to V-Net for the segmentation task.  


Attention mechanisms have been proved quite efficient in learning better feature representations. \cite{Liu2020pyramid} proposed a feature pyramid attention sub-module before the decoder in FCN, considering the semantic information from U-Net may not be sufficient to represent the heterogeneous anatomic structures for a clear boundary. \cite{ding2023multi} interleave the U-Net skip connections with a multi-scale self-attention mechanism for recalibrating the feature maps across multiple layers. \cite{wang2023two} developed a two-stage approach, where they employ a Squeeze and Excitation (SE) CNN for detecting the prostate's existence in stage-1 and a Residual-Attention U-Net in stage-2 for segmenting the slices that include the prostate gland. \cite{li2023dual} propose a dual attention mechanism using 3D convolutions to learn in an end-to-end manner both the gland and lesion segmentation tasks.


An interesting work of \cite{jia20193d} proposes 3D APA-Net, a 3D adversarial pyramid anisotropic convolutional deep neural network for prostate segmentation, which has an encoder-decoder architecture, equipped with adversarial training for spatially consistent and continuous segmentation results.  Recently, a novel attention mechanism among slices is proposed from \cite{hung2022cat} to learn cross-slices features at multiple scales using transformer blocks. A more generic approach of \cite{chen2023semi} proposes a semi-supervised learning method for segmenting medical images, using attention among slices to capture the common spatial layout of patients organs and a contrastive learning scheme to incorporate unlabeled data in training. 


In some works, certain modules are proposed to help make the boundary more clear and accurate. For example, \cite{jia2019hd} proposed hybrid discriminative network named HD-Net, in which the decoder consists of two branches: a 3D segmentation branch and a 2D boundary branch to boost the shared encoder to learn features with more semantic discrimination. \cite{zhu2019boundary} proposed a BOWDA-Net where a boundary-weighted segmentation loss was introduced to the transfer learning.


\subsection{Successive subspace learning methodology}

Recently, being inspired by deep learning but being different, the
successive subspace learning (SSL) methodology was proposed by Kuo {\em
et al.} in a sequence of papers \cite{kuo2016understanding, kuo2017cnn,kuo2019interpretable}. Instead of using back-propagation,
feature representations in SSL-based methods are learnt in an
unsupervised feedforward manner using multi-stage principal component
analysis (PCA) for multi-scale subspace learning. The overall framework
is named Green Learning, as it is meant to offer image analysis and
understanding solutions with much less parameters and complexity. 

Three variants of transforms were originally proposed, including Saak
(subspace approximation with augmented kernels)
transform~\cite{kuo2018data}, the Saab (subspace
approximation via adjusted bias) transform~\cite{kuo2019interpretable}, and the channel-wise (c/w) Saab transform~\cite{chen2020pixelhop++}.
Among them, the c/w Saab transform requires the smallest model size and has the best transform efficiency because it takes advantage of the weak correlation between channels so that filters are learnt from each channel separately. The details of c/w Saab transform is introduced in Section~\ref{subsec:encoder}. 

SSL methodology has been applied to many domains in image processing and
computer vision.
Particularly, two representative works of SSL-based method in medical
image analysis can be found in~\cite{liu2021voxelhop} and
\cite{liu2021segmentation}, which solve ALS disease classification and
cardiac MRI segmentation, respectively. Our proposed PSHop can find the
closest shadows of~\cite{liu2021segmentation} and is the first work that
generalizes the SSL methodology on a 3D medical image segmentation.

\section{Methods}\label{sec:methods}

\begin{figure*}[t]
\begin{center}
\includegraphics[width=0.9\linewidth]{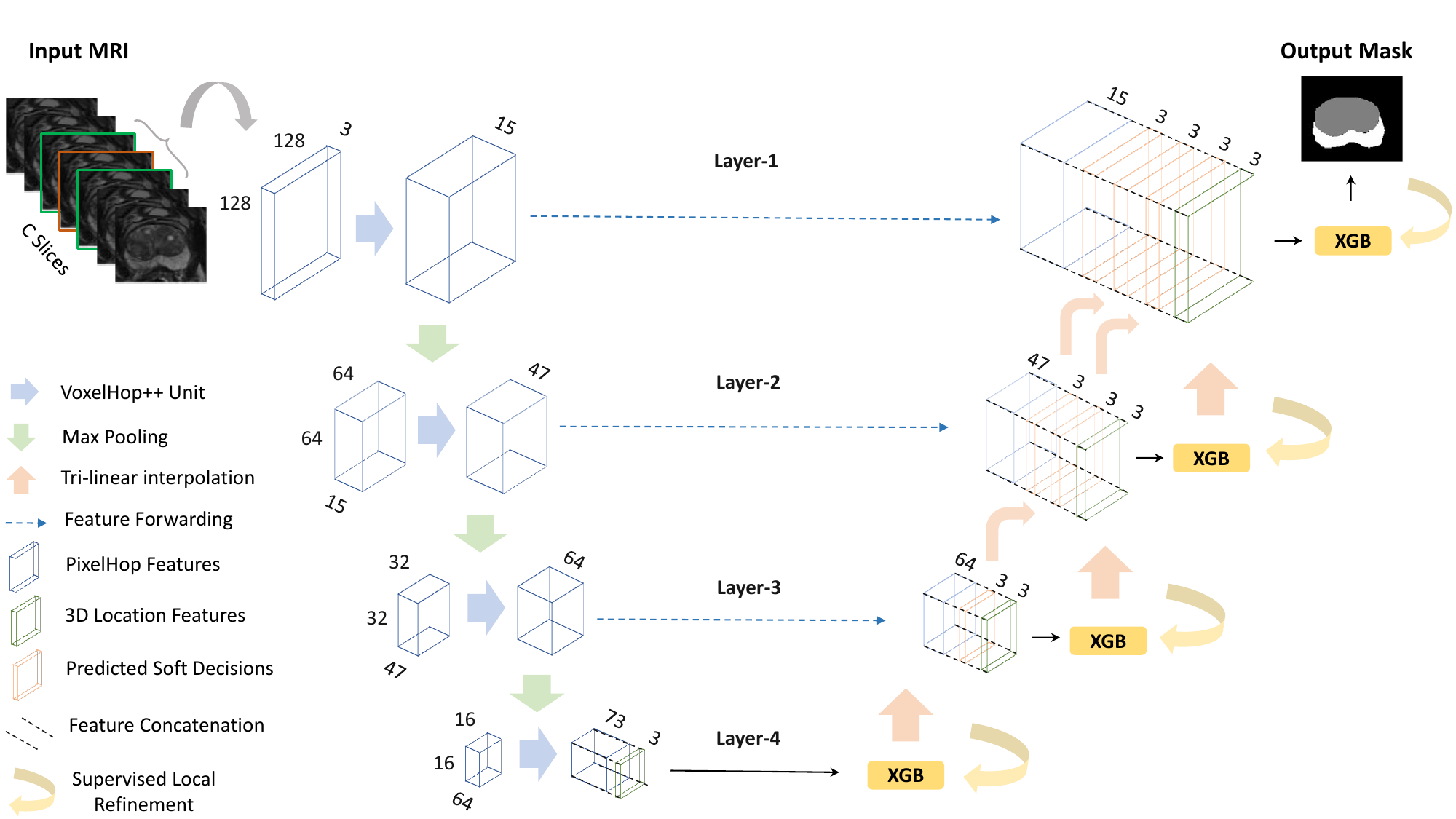}
\end{center}
\caption{Illustration of the U-Net-like 3D architecture of PSHop for
representation learning and feature extraction. Four scales (layers) are
used in PSHop. Deeper layers correspond to coarser features, while
shallower ones are meant to refine the segmentation result and output
the segmentation mask at the input's scale. The process is shown for
segmenting one slice from the entire sequence. It is
repeated for every single slice in the MRI input.}
\label{fig:overall_pipeline}
\end{figure*}

A complete solution to prostate segmentation entails two tasks: (1) the
whole gland segmentation and (2) the zonal segmentation. Each of those
tasks has its importance within the prostate medical diagnosis pipeline. Our proposed PSHop method is used for
both tasks separately and its architecture overview is shown in
Figure~\ref{fig:overall_pipeline}, which has an encoder-decoder
structure, inspired by U-Net structure. In the encoder part,
unsupervised feature representations of different scales are extracted
from fine to coarse. In the decoder part, segmentation mask is predicted
in a coarse to fine fashion based on the representations from the
encoder. The details of the method are described in
Sec.~\ref{subsec:encoder} and Sec.~\ref{subsec:decoder}. 

\subsection{Encoder: Fine to Coarse Representation Learning}\label{subsec:encoder}

We treat the segmentation task as a voxel-wise classification problem
and extract the representation of a neighborhood for each voxel in
different scales. Different from deep learning based methods where the
convolutional filters are learnt through end-to-end optimization of the
loss function, we use cascaded VoxelHop units~\cite{liu2021voxelhop}
which is a statistical approach to extract feature vectors. The process
is fully unsupervised, with no ground truth labels required during
training. It is also a feed-forward learning process instead of using
back-propagation in deep learning based encoders.

\subsubsection{Feature Extraction based on VoxelHop Units}
Figure~\ref{fig:Skip_connection_HOP} illustrates the process of each
VoxelHop unit. It consists of two consecutive steps: 1) neighborhood
construction in the 3D space, and 2) representation learning through the
c/w Saab transform~\cite{chen2020pixelhop++}. 

Suppose the input tensor of the $i$-th VoxelHop unit is of dimension
$H_i\times W_i\times C_i \times K_i$, where $H_i$, $W_i$ and $C_i$
represent the resolution in 3D space, and $K_i$ represents the
dimension of the feature vector for each voxel extracted from the
$(i-1)$-th VoxelHop unit. Specifically, for the first VoxelHop unit
where $i=1$, $H_1\times W_1\times C_1$ corresponds to the input MRI data
resolution, and $K_1=1$. We first gather the neighborhood in the 3D
space centered at each voxel. The neighborhood size is defined as
$S_{Hi}\times S_{Wi}\times S_{Ci}$ in spatial. Each voxel in the
neighborhood has a feature vector of dimension $K_i$, which results in a
tensor of size $S_{Hi}\times S_{Wi}\times S_{Ci}\times K_i$.

The neighborhood tensor is then flattened in the spatial domain. Channel-wise Saab transform is performed in each of the $C_i$ channels separately to learn spectral signals through subspace approximation at the current scale. Suppose the input vector is $\mathbf{x}\epsilon \mathbb{R}^{N}$, where $N=S_{Hi}\times S_{Wi} \times S_{Ci}$. Features can be extracted by projecting the input vector on to several anchor vectors, which can be expressed as an affine transform expressed as
\begin{equation}\label{eq:affine_transform}
y_m = \mathbf{a}_m^T\cdot \mathbf{x} + b_m, m=0,1,\cdots, M-1,
\end{equation}
where $\mathbf{a}_m$ is the $m$-th anchor vector of dimension $N$, and $M$ is the total number of anchor vectors. Here, the channel-wise Saab transform is a data-driven approach to learn the anchor vectors from all the neighborhood tensors collected from the input data. First, it decomposes the input subspaces into the direct sum of two subspaces, i.e. DC and AC, expressed in Eq.~\ref{eq:SDC_SAC}, where the terms are borrowed from the ``direct circuit'' and ``alternating circuit'' in the circuit theory.
\begin{equation}\label{eq:SDC_SAC}
S = S_{DC} \oplus S_{AC}.
\end{equation}
$S_{DC}$ and $S_{AC}$ are spanned by $DC$ and $AC$ anchor vectors, defined as:
\begin{itemize}
\item DC anchor vector $\mathbf{a}_0=\frac{1}{\sqrt{N}}\left ( 1, 1, \cdots , 1 \right )^T$
\item AC anchor vectors $\mathbf{a}_m$, $m=1,\cdots, M-1$.
\end{itemize}
The two subspaces are orthogonal to each other, where the input signal
$\mathbf{x}$ is projected on to $\mathbf{a}_0$ to get the DC component
$\mathbf{x}_{DC}$. Then the AC component is extracted by subtracting DC
component from the input signal, i.e.
$\mathbf{x}_{AC}=\mathbf{x}-\mathbf{x}_{DC}$. 

\begin{figure}[t]
\begin{center}
\includegraphics[width=1.0\linewidth]{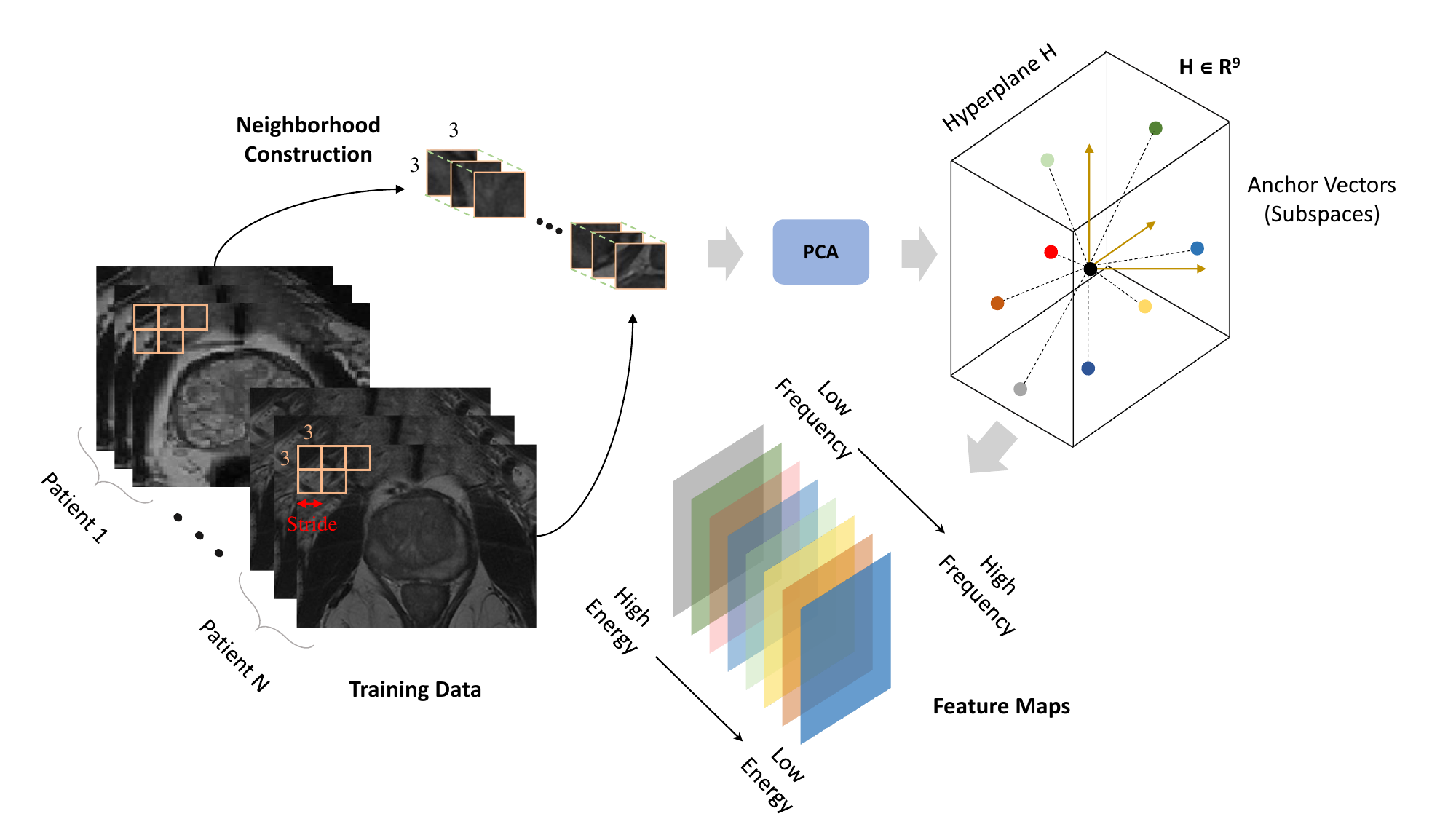}
\end{center}
\caption{An illustration of the local neighborhood construction for unsupervised filter learning using the Saab transform based on PCA. Each anchor vector corresponds to a subspace on the N-dimensional plane (denoted in different colors). The input image is projected on to these subspaces to obtain its spectral decomposition at a certain scale that corresponds to the output feature map. In c/w Saab this decomposition is applied on every single feature map in a recursive manner, until the maximum number of layers is reached.}
\label{fig:voxelhop_unit}
\end{figure}

After that, AC anchor vectors are learnt by conducting principal
component analysis (PCA) on the AC component. The first $K$ principal
components are kept as the AC anchor vectors. Thus, one can extract
features by projecting $\mathbf{x}$ on to the above learnt anchor vectors
based on Eq.~\ref{eq:affine_transform}. The bias term is selected to
ensure all features are positive by
following~\cite{kuo2019interpretable}. An illustration of the feature
extraction concept behind the VoxelHop unit using the derived subspaces
from PCA is provided in Fig. \ref{fig:voxelhop_unit}. 

\subsubsection{Representation Learning Through Cascaded VoxelHop Units}

Each of the above presented VoxelHop unit extracts the representation in
a certain resolution. For segmentation tasks, both local and global
descriptions are important. Features that represent a small neighborhood
serve for better localization, while features from a larger neighborhood
provide a more accurate context semantic understanding. 

\begin{figure*}[t]
\begin{center}
\includegraphics[width=0.8\linewidth]{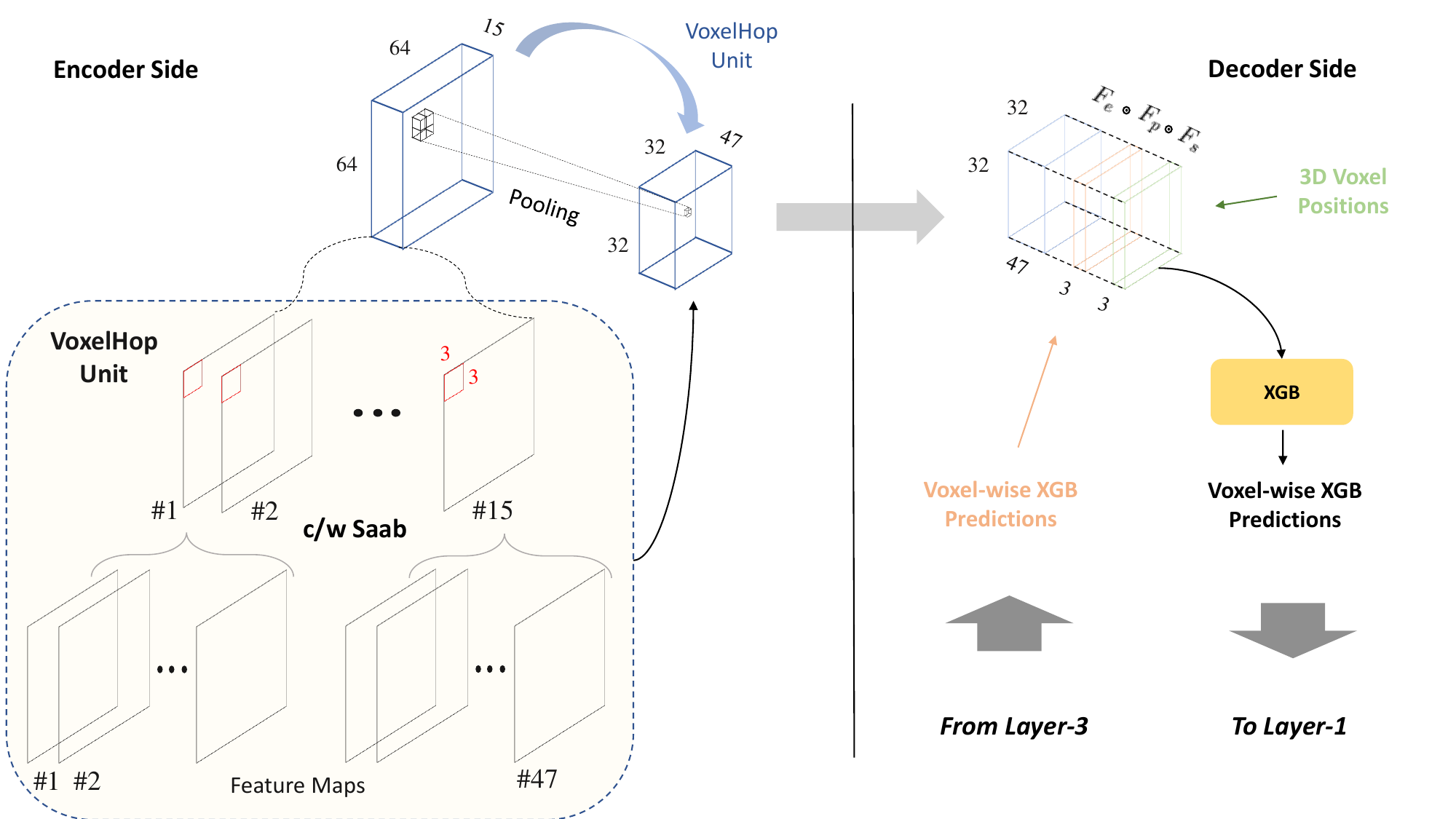}
\end{center}
\caption{Feature representation in one VoxelHop layer within PSHop. Layer 2 is borrowed as example to show the  connection between encoder-decoder and feature concatenation from the encoder $F_e$, the voxel-wise probabilities from the coarser scale after interpolation $F_p$ and the 3D voxel positions $F_s$. The concatenated feature is fed in XGB classifier to predict the voxel-wise features for layer-1.}
\label{fig:Skip_connection_HOP}
\end{figure*}

In the PSHop encoder, we use cascaded VoxelHop units with max-pooling in
between. Thus, the receptive field of the representations grows fast as
more VoxelHop units are cascaded until a preset number of cascaded
layers, $L$, is reached. The detailed architecture of the proposed PSHop
encoder is summarized in Table~\ref{tab:encoder_archi}. We set $L=4$,
where VoxelHop units from Hop-1 to Hop-4 form a representation learning
from fine to coarse. Here, we use ``hop'' to represent a certain
neighborhood range. The four cascaded hops can be constructed as a
tree-decomposed structure, where the $i$-the VoxelHop unit yields the
$i$-th child nodes (see Fig.~\ref{fig:Skip_connection_HOP}). The number
of Saab filters for each hop unit is decided by the input data using a
pre-set energy threshold, where the energy of a child node refers to the
multiplication of the energy of the parent node and the normalized
energy from PCA among child nodes in the same level. 

\begin{table}[t]
\centering
\caption{Encoder Architecture of the Proposed PSHop}
\label{tab:encoder_archi}
\begin{tabular}{llll}
\toprule
  & Filter Size &  Stride \\
\midrule
VoxelHop 1        & $(3\times3)\times3$&  $(1\times1)\times1$& \\
Max-pooling 1     & $(2\times2)\times2$&  $(2\times2)\times2$& \\
VoxelHop 2        & $(3\times3)\times3$&  $(1\times1)\times1$& \\
Max-pooling 2     & $(2\times2)\times2$&  $(2\times2)\times2$& \\
VoxelHop 3        & $(3\times3)\times3$&  $(1\times1)\times1$& \\
Max-pooling 3     & $(2\times2)\times2$&  $(2\times2)\times2$& \\
VoxelHop 4        & $(3\times3)\times3$&  $(1\times1)\times1$& \\ 
\bottomrule
\end{tabular}
\end{table}

\subsection{Decoder: Coarse to Fine Segmentation Prediction}\label{subsec:decoder}

The segmentation is conducted from coarse to fine based on the features from PSHop encoder. We first start from the deepest hop, Hop-$L$, and perform a one-level PSHop decoder unit. Then the process gradually move to a shallower hop unit until it outputs the segmentation prediction for the full input resolution. Each PSHop decoder unit consists of the following three steps: 1) feature aggregation; 2) segmentation and local refinement at the current scale; and 3) prediction upsampling.
\subsubsection{Feature Aggregation}
The feature for the segmentation at each scale comes from different sources. Besides the \textit{encoder features} $F^i_e$ at the $i$-th hop, a \textit{position encoding} feature vector $F^i_s$ is included where the 3D voxel coordinate is recorded, expressed as
\begin{equation}\label{eq:F_s}
F^i_{s} = \left [ x,y,z \right ]^T. 
\end{equation}
Also, the \textit{predicted probability} vectors $F^i_p$ upsampled from all the coarser hops is also included. The final feature vector for segmentation prediction is an aggregation of the above mentioned three features detailed as
\begin{equation}\label{eq:F_final}
F^i_{seg} = F^i_s \odot F^i_p \odot F^i_e, 
\end{equation}
where $F^i_e$ is copied from encoder to the decoder at the $i$-th hop using a skip connection, and $\odot$ represents the voxel-wise concatenation operation. A detailed illustration can be found in Figure \ref{fig:Skip_connection_HOP}.

The three parts serve as different roles. First, the relative physical positions of different structures is similar among different patients in MRI images, for example, the prostate is always around the center region in the 2D plains. The position encoding $F_s$ helps merge this prior knowledge into the prediction process. Second, $F^i_p$ is propagated from all the coarser grids, which contains the probability of classes predicted using different receptive fields that are larger than that of the $i$th hop. Thus, $F^i_p$ provides a memory of coarse to fine context semantic information. With these conditions, the role of $F^i_e$ is to provide a representation of the local neighborhood of each voxel so that the segmentation prediction can result in a finer detail.

\subsubsection{Segmentation Prediction and Local Refinement}
We treat the segmentation at each scale as a voxel-wise classification problem. In the training process, the ground truth masks are first encoded as one-hot vectors representing the corresponding class. The ground truth of each voxel grid at the $i$th hop is then downsampled from ground truth in the original resolution using bilinear interpolation. The selection of the training samples is based on the confidence level of the interpolated ground truth mask. Only the voxels with high confidence is included as the training samples. Then, an eXtreme Gradient Boosting (XGBoost~\cite{xgb}) classifier is trained using the aggregated feature $F^i_{seg}$. The output is the predicted soft decision vectors at the current scale. 

Since the segmentation requires smoothness in a local neighborhood while the classification solution is made for each voxel separately, we adopt a local refinement step after each XGBoost classifier based on the soft decisions. Inspired by the soft-label smoothing (SLS) technique proposed in~\cite{yang2021pixelhop}, we gather a cubic of soft decisions in the 3D domain of size $3\times 3\times 3$. The aggregated neighborhood soft decisions are concatenated as the new feature to train another XGBoost classifier. This is repeated iteratively. In practice, we perform two iterations for the soft decision update. In this way, the resulted segmentation prediction is smoothed and the precision is improved. This is used at all scales of the prediction, up to layer-1 which gives the final segmentation output. To further refine the output and correct any segmentation artifacts, we employ a median filter of size $7\times7$ as a  post-processing step. In the experimental section we demonstrate the effectiveness of the post-processing filter.

\subsubsection{Prediction Upsampling}
To propagate the predicted soft decisions of a coarser grid to a finer grid, we perform a bilinear upsampling towards the target resolution. This process is cumulative, which means that the predicted probability vector $F^i_p$ propagated to the $i$-th hop is from all the coarser hops, detailed as
\begin{equation}\label{eq:F_p}
F^i_{p} = \left \{ F^{i+1}_p \odot \hat{y}^{i+1} \right \} \uparrow , \quad
\forall \hspace{1mm} i \leq L-1,
\end{equation}
where $\hat{y}^{i+1}$ is the predicted soft decision at the $(i+1)$-th hop, and $\left \{ \cdot \right \}\uparrow$ means the bilinear upsampling.

\section{Experimental Setup}\label{sec:experiments_setup}
\subsection{Database and pre-processing}

To demonstrate the effectiveness of the proposed PSHop method, we conduct experiments based on one public MR image database NCI-ISBI 2013 Challenge (Automated Segmentation of Prostate Structures \cite{isbi}) and one private in-house USC-Keck dataset. The ISBI-2013 dataset consists of 60 training cases of axial T2-weighted MR 3D series, where half were obtained at 1.5T (Philips Achieva at Boston Medical Center) and the other half at 3T (Siemens TIM at Radboud University Nijmegen Medical Center). Since the ground truth segmentation includes background, peripheral zone (PZ), and transitional zone (TZ), we merge PZ and TZ as one class -- prostate area, for the model training and evaluation since we consider the prostate segmentation task in this paper. The pixel spacing within each slice ranges from 0.39 \textit{mm} to 0.75 \textit{mm}, while the through-plane resolution ranges from 3.0 \textit{mm} to 4.0 \textit{mm} among different patients. 

The USC-Keck dataset consists of a cohort of 260 patients collected in the Keck Medicine School in the University of Southern California. Besides T2-w 3D series, for each patient T2-Cube series is also available. It is known that T2-Cube has smaller pixel spacing, especially along the z-axis. That means, higher resolution and thinner slices. Specifically, for T2-w pixel spacing ranges from 0.5 \textit{mm} to 0.7 \textit{mm} and the through-plane resolution (z-axis) from 3.0 mm \textit{mm} to 4.0 \textit{mm}. For T2-Cube pixel spacing is set at 0.83 \textit{mm} and the through-plane resolution (z-axis) at 1.4 \textit{mm}. The scanner used to acquire those images is the GE 3T with 8ch Cardiac coil. In our experimental analysis with USC-Keck data, the T2-Cube series is used since it can presumably provide more accurate segmentation results because of the higher perspicuity of the images (stemming from the smaller voxel spacing).

For both datasets, we first regularize the resolution of different images to the same physical resolution of $0.625\times 0.625 \times 1.5 \, mm^3$. We use Lanczos interpolation where the factor is calculated based on the original pixel spacing and through-plane resolution of each image. Here, the through-plane resolution is increased so that the segmentation in the 3D space can be more accurate. After that, to reduce the artifacts while acquiring the images, contrast enhancement using CLAHE~\cite{CLAHE} is applied. Finally, before feeding PSHop input, for the whole gland segmentation task the input sequence is resized to $128\times128$. For the zonal segmentation task, a $256\times256$ centered crop around the segmented gland is resized to $128\times128$ to standardize the PSHop input.  

\subsection{Evaluation Metrics}
To quantitatively evaluate the performance, we calculate the Dice Similarity Coefficient (DSC) \cite{Dice1945MeasuresOT} expressed in Eq. (\ref{eq:dsc}) in a binary scenario, where $X$ and $Y$ represent the ground truth and the predicted segmentation mask, respectively. DSC is widely used in evaluating segmentation tasks for medical images. It measures the ratio of the intersection of two binary sets to the averaged cardinality. 
\begin{equation}
    DSC\left (X,Y\right ) = \frac{2\left | X\cap Y \right |}{\left | X \right|+\left | Y \right |}.
    \label{eq:dsc}
\end{equation}



\subsection{Experimental Details}
For both datasets, we apply the same experimental settings. For ISBI-2013 and USC-Keck we apply 5-fold cross validation on the 60 and 260 training images, respectively and calculate the mean and standard deviation of the evaluation scores. That is, for ISBI-2013 $48$ sequences are used for training PSHop from each fold and the rest $12$ for validation. For USC-Keck dataset $206$ sequences are used for training and the rest $54$ for validation. 

For benchmarking PSHop with other DL-based methods we conduct the same experiment setting for both datasets, and train V-Net and U-Net based architectures. Besides segmentation performance comparison using the DSC metric, we also compare the model size and complexity of the models, since the main target and motivation of this work is to offer a lightweight solution comparing to other methods.



\section{Experimental Results}\label{sec:results}
\subsection{Segmentation Results}
The DSC scores of ISBI-2013 and USC Keck datasets obtained by our proposed PSHop method for the prostate gland segmentation task are summarized in Table~\ref{tab:dsc_whole_benchmark_ISBI}. The averaged validation performance over the 5-folds is reported and the standard deviation as well (shown in parenthesis). To begin with, PSHop has a very competitive performance among the two baseline DL-based works. In particular, PSHop outperfoms V-Net on ISBI-2013 dataset and U-Net on the USC-Keck one. ISBI-2013 has considerably fewer patient data than USC-Keck. Therefore, V-Net performs better when it is given with sufficient training patients, while U-Net has a higher performance with fewer training samples. That is also evident from the high standard deviation on ISBI data from V-Net. Another important observation is that PSHop has a more stable performance (low standard deviation in both experiments), regardless the number of training samples. This underlines one of the GL framework advantages that is more stable even for fewer training samples, while large DL models fail to achieve a high performance when data are scarce. That also confirms GL's main assumption that statistical-based feed-forward models can still perform well even with a small number of training samples, which is usually the case for medical imaging datasets.

Steering our comparisons to the zonal segmentation performance, Table \ref{tab:dsc_USC_benchmark_zonal} shows the benchmarking on USC-Keck dataset, since it is fairly larger than the ISBI-2013 and thereby stronger conclusions can be drawn. For TZ, PSHop surpasses U-Net by large margins and maintains a small performance gap with V-Net. On the other hand, one can observe that the performance gap is larger on the PZ when compared with V-Net and this is an area for further improvement for GL. Yet, PSHop achieves a much higher DSC score comparing to the U-Net, which is the baseline architecture for many works throughout the literature. For the smaller ISBI-2013 dataset PSHop outperforms the other methods on the PZ (see Table \ref{tab:dsc_ISBI_benchmark_zonal}), mainly due to the small number of training samples that PSHop has an advantage. On the TZ, PSHop surpasses the U-Net performance by large margins.

In the above comparisons, V-Net generally performs better when trained with a large number of samples and U-Net performs well only on the ISBI-2013, which has many fewer samples. This is sensible because V-Net has many more trainable parameters comparing to U-Net and thus can fit better the data diversity. Table \ref{tab:complexity_benchmark} demonstrates the green advantages and benefits of GL when it comes to complexity and model size comparison. PSHop has an order of magnitude less parameters than the DL-based models. Also, in terms of complexity, it has $\times$190 less FLOPS than U-Net and $\times$5269 than V-Net. These comparisons stress the tremendous advantages of GL-based solutions for application deployment.  

Overall, PSHop maintains a very competitive standing performance-wise with other DL baseline models, outperforming in general U-Net in both tasks and having a close performance with V-Net. Nevertheless, V-Net's performance comes at the expense of a higher complexity and model size. 

 
\begin{table}[t]
\centering
\caption{Comparison of the whole gland segmentation performance with PSHop and two popular baseline deep learning models using the DSC metric.}
\label{tab:dsc_whole_benchmark_ISBI}
\begin{tabular}{lcc}
\toprule
       & ISBI-2013   & USC-Keck \\ 
\midrule
V-Net  & 0.762 ($\pm$0.139) & 0.906 ($\pm$ 0.009)     \\ 
2D U-Net & 0.684 ($\pm$0.031) &  0.809 ($\pm$ 0.036)   \\ 
\midrule
\textbf{PSHop (Ours)}   & 0.826 ($\pm$0.018) & 0.873 ($\pm$ 0.017)    \\ 
\bottomrule
\end{tabular}
\end{table}

\begin{table}[t]
\centering
\caption{Comparison of the zonal segmentation performance with PSHop and two popular baseline deep learning models using the DSC metric on USC-Keck data.}
\label{tab:dsc_USC_benchmark_zonal}
\begin{tabular}{lcc}
\toprule
       & TZ & PZ \\ 
\midrule
V-Net  &  0.878 ($\pm$ 0.019) & 0.747 ($\pm$ 0.014)     \\ 
U-Net & 0.741 ($\pm$ 0.041) & 0.525 ($\pm$ 0.032)   \\ 
\midrule
\textbf{PSHop (Ours)}  & 0.845 ($\pm$ 0.025) & 0.656 ($\pm$ 0.012)     \\ 
\bottomrule
\end{tabular}
\end{table}

\begin{table}[t]
\centering
\caption{Comparison of the zonal segmentation performance with PSHop and two popular baseline deep learning models using the DSC metric on ISBI-2013 data.}
\label{tab:dsc_ISBI_benchmark_zonal}
\begin{tabular}{lcc}
\toprule
       & TZ & PZ \\ 
\midrule
V-Net  &  0.629 ($\pm$ 0.064) & 0.431 ($\pm$ 0.030)     \\ 
2D U-Net & 0.563 ($\pm$ 0.020) & 0.316 ($\pm$ 0.031)   \\ 
\midrule
\textbf{PSHop (Ours)}  & 0.667 ($\pm$ 0.026) & 0.366 ($\pm$ 0.039)     \\ 
\bottomrule
\end{tabular}
\end{table}


\begin{figure*}[t]
\begin{center}
\includegraphics[width=1.0\linewidth]{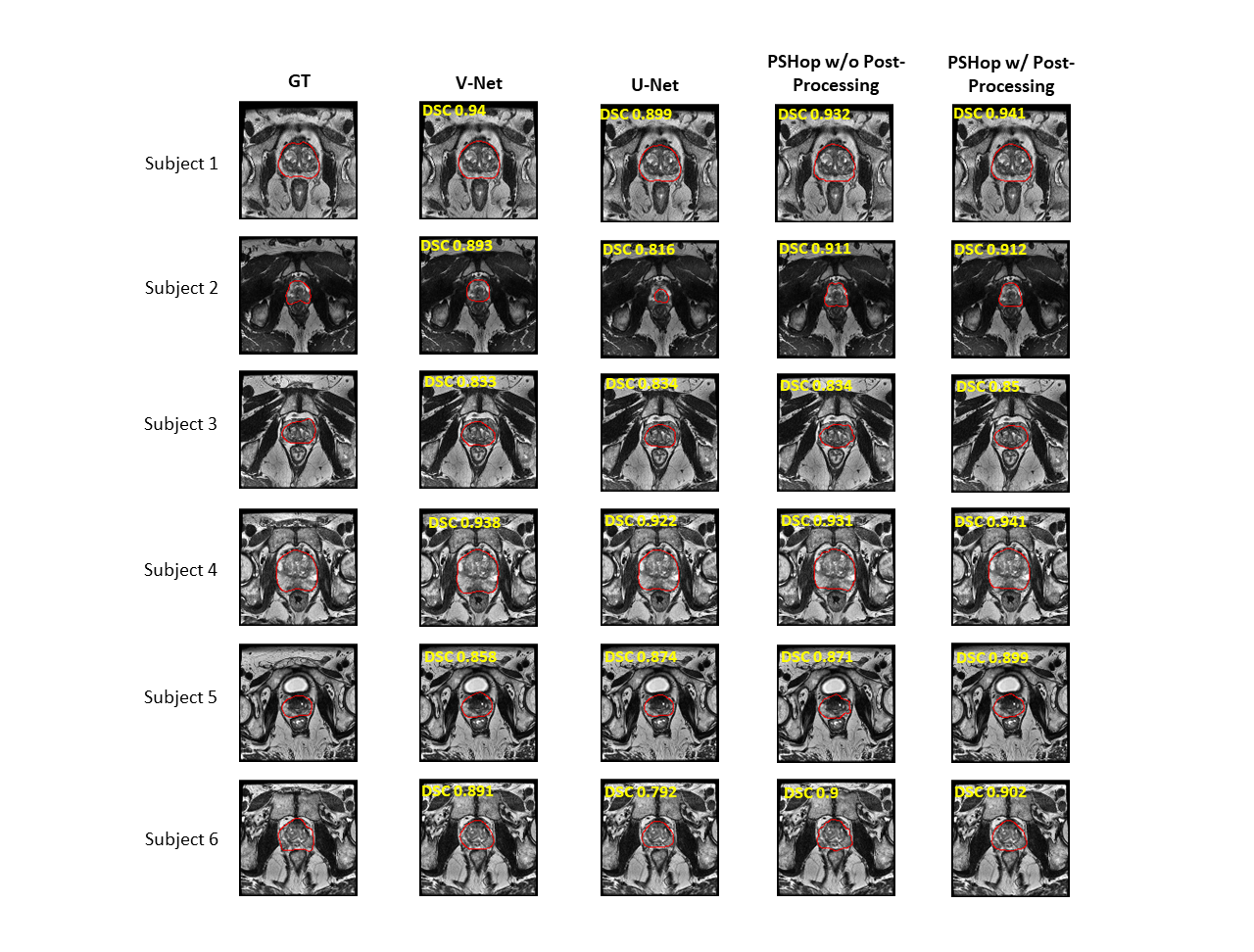}
\caption{Qualitative comparison among DL methods and PSHop on the whole gland segmentation task. Results are also shown before and after the post-processing operation of the median filter.} \label{fig:vis_whole}
\end{center}
\end{figure*}

\begin{figure*}[t]
\begin{center}
\includegraphics[width=0.85\linewidth]{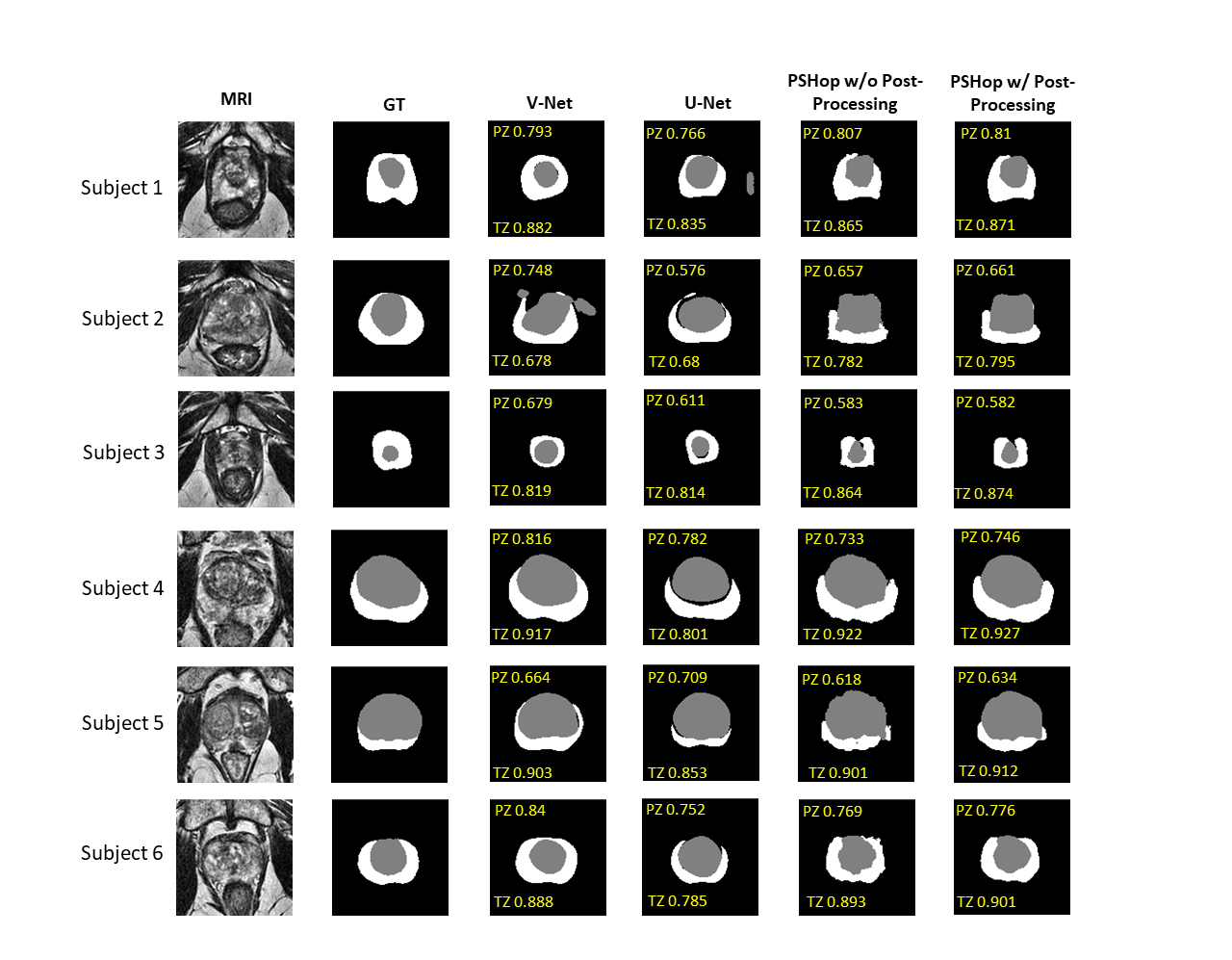}
\caption{Qualitative comparison among DL methods and PSHop on the zonal segmentation task.} \label{fig:vis_zonal}
\end{center}
\end{figure*}

\begin{table}[t]
\centering
\caption{Comparison of model size and complexity in inference between PSHOP and the two deep learning baseline models.}
\label{tab:complexity_benchmark}
\begin{tabular}{lcc}
\toprule
       & \# of parameters & FLOPS \\ 
\midrule
V-Net  & 45,603,934 & 379B ($\times 5269$) \\ 
2D U-Net & 17,970,626 & 13.6B ($\times 190$)\\ 
\midrule
\textbf{PSHop (Ours)} & 235,206 & 72M ($\times 1$)\\ 
\bottomrule
\end{tabular}
\end{table}

In Figures \ref{fig:vis_whole} and \ref{fig:vis_zonal} we provide qualitative comparisons for the whole gland and zonal segmentation, respectively. One observation is the segmentation refinement from the post-processing module for both tasks. Moreover, our module performs better on segmenting the TZ than PZ which is usually a more challenging task for various models, due to its irregular shape variations along slices.

\section{Conclusion}\label{sec:conclusion}
This work proposes the PSHop method for accurate prostate gland and zonal segmentation. Unlike other majority state-of-the-art works based on DNNs, such as U-Net or V-Net, PSHop follows the GL paradigm and adopts a feed-forward model for feature extraction. The model architecture is inspired by U-Net, where there are several multi-scale representations of the input MRI, nevertheless PSHop uses no back-propagation, but the SSL methodology to extract feature representations. PSHop has a competitive performance standing among other DL-based methods, outperforming U-Net on the larger dataset and has a small performance margin with V-Net. All in all, PSHop method comes with a very lightweight model size and orders of magnitude less computational complexity, thus providing a green alternative for prostate segmentation task. Additionally, the linear feature extraction model provides more transparency in the pipeline and hence makes PSHop decisions more trustworthy to physicians.

\printcredits

\bibliographystyle{model2-names}

\bibliography{cas-refs}



\end{document}